\documentclass{article}

    \PassOptionsToPackage{numbers, compress}{natbib}

\usepackage[final]{neurips_2023}

\usepackage[utf8]{inputenc} 
\usepackage[T1]{fontenc}    
\usepackage{hyperref}       
\usepackage{url}            
\usepackage{booktabs}       
\usepackage{amsfonts}       
\usepackage{nicefrac}       
\usepackage{microtype}      
\usepackage{multirow}

\usepackage{amsmath,amsfonts,bm}

\def\eqref#1{equation~\ref{#1}}

\def\1{\bm{1}}

\def\rva{{\mathbf{a}}}

\def\rvv{{\mathbf{v}}}

\def\rmI{{\mathbf{I}}}

\DeclareMathAlphabet{\mathsfit}{\encodingdefault}{\sfdefault}{m}{sl}
\SetMathAlphabet{\mathsfit}{bold}{\encodingdefault}{\sfdefault}{bx}{n}

\def\gN{{\mathcal{N}}}

\newcommand{\Ls}{\mathcal{L}}
\newcommand{\R}{\mathbb{R}}

\newcommand{\raw}{\mathrm{raw}}
\newcommand{\mask}{\texttt{[MASK]}}

\usepackage[dvipsnames]{xcolor}
\usepackage{graphicx}
\usepackage{wrapfig}
\usepackage{enumitem}
\usepackage[textwidth=2.5cm]{todonotes}

\usepackage{amsmath}
\usepackage{amssymb}
\usepackage{makecell}
\usepackage[capitalize]{cleveref}

\usepackage{pifont}
\newcommand{\cmark}{\ding{51}}%
\newcommand{\xmark}{\textcolor{lightgray}{\ding{55}}}%
\usepackage{array}
\newcolumntype{L}{>{\centering\arraybackslash}m{2cm}}

\title{Diffusion Models as Masked Audio-Video Learners}

\author{%
  Elvis Nunez\thanks{Work done during an internship at Apple.} \\
  University of California, Los Angeles\\
  \texttt{elvis.nunez@ucla.edu} \\
  \And
  Yanzi Jin \\
  Apple \\
  \texttt{yanzi\_jin@apple.com} \\
  \AND
  Mohammad Rastegari \\
  Apple \\
  \texttt{mrastegari@apple.com} \\
  \And
  Sachin Mehta \\
  Apple \\
  \texttt{sachin\_mehta@apple.com} \\
  \And
  Maxwell Horton \\
  Apple \\
  \texttt{mchorton@apple.com} \\
}

\begin{document}

\maketitle

\begin{abstract}
Over the past several years, the synchronization between audio and visual signals has been leveraged to learn richer audio-visual representations. Aided by the large availability of unlabeled videos, many unsupervised training frameworks have demonstrated impressive results in various downstream audio and video tasks. Recently, Masked Audio-Video Learners (MAViL) has emerged as a state-of-the-art audio-video pre-training framework. MAViL couples contrastive learning with masked autoencoding to jointly reconstruct audio spectrograms and video frames by fusing information from both modalities. In this paper, we study the potential synergy between diffusion models and MAViL, seeking to derive mutual benefits from these two frameworks. The incorporation of diffusion into MAViL, combined with various training efficiency methodologies that include the utilization of a masking ratio curriculum and adaptive batch sizing, results in a notable 32\% reduction in pre-training Floating-Point Operations (FLOPS) and an 18\% decrease in pre-training wall clock time. Crucially, this enhanced efficiency does not compromise the model's performance in downstream audio-classification tasks when compared to MAViL's performance.
\end{abstract}

\section{Introduction}
\begin{figure}[t!]
    \centering
   \includegraphics[width=\linewidth]{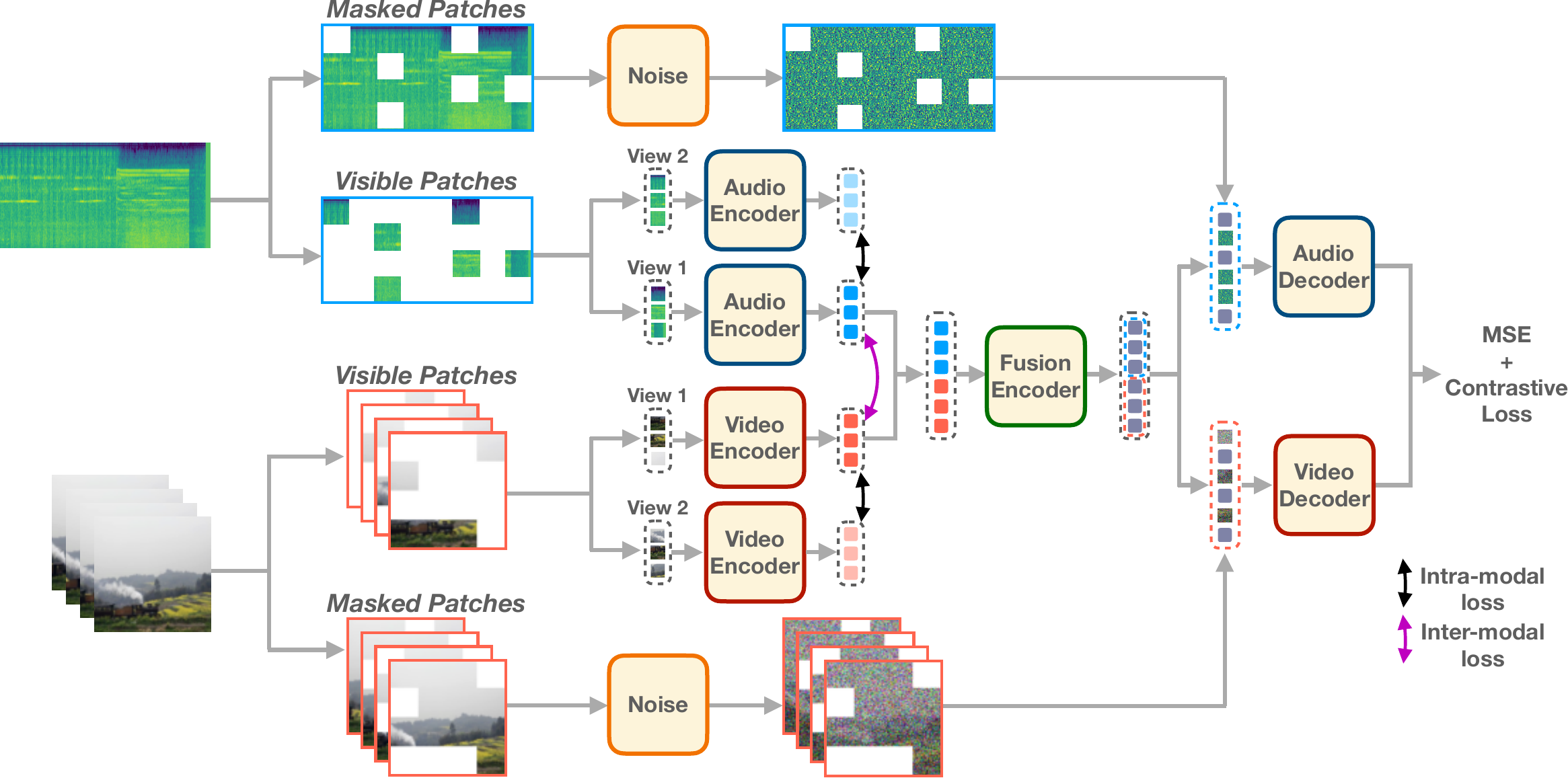}
   \caption{\textbf{DiffMAViL architecture.} Similar to the audio-video encoder-decoder architecture of MAViL \cite{MAViL}, our DiffMAViL architecture takes as input RGB video frames and audio spectrograms. The spectrogram and RGB frames are first randomly masked, and visible patches from each modality are encoded via their respective encoders. Masked patches are diffused and concatenated with the outputs of the audio-video fusion encoder, which are then fed through the audio and video decoders to obtain reconstructions of the input spectrogram and RGB frames.}
\label{fig:diffmavil}
\end{figure}

Large-scale unsupervised pre-training has improved the accuracy of down-stream tasks in the audio-visual domain \cite{MAViL, CavMAE, soundnet, korbar2018cooperative, looklistenlearn}. A common approach to self-supervised learning involves specifying a pre-text task \cite{doersch2015unsupervised, wang2015unsupervised, noroozi2016unsupervised, gidaris2018unsupervised}, whereby supervisory signals are extracted from large amounts of unlabeled data in an effort to facilitate the learning of meaningful representations. For example, denoising autoencoders \cite{denoisingautoencoder} aim to learn representations by reconstructing input samples from noisy data. More recently, masked autoencoders (MAE) \cite{ImageMAE, SpatiotemporalMAE, VideoMAE, VideoMAEv2, AudioMAE, DiffusionMAE}, aim to learn representations by randomly masking large portions of the input and attempting to reconstruct the original input via a mean-squared-error minimization. This simple approach has demonstrated strong performance across different modalities, including image \cite{ImageMAE}, audio \cite{AudioMAE}, and video \cite{SpatiotemporalMAE, VideoMAE, VideoMAEv2}. Moreover, several works have explored multi-modal frameworks, combining audio and video domains \cite{MAViL, CavMAE}. In an effort to facilitate the learning of high-frequency features, the MAE framework has also been cast in the context of diffusion models \cite{DiffusionMAE}, whereby reconstructions are shown to exhibit higher frequency details. While self-supervised pre-training has witnessed great success in various downstream tasks, pre-training remains a computationally expensive procedure, often requiring thousands of GPU hours \cite{rombach2022high, dhariwal2021diffusion, chowdhery2022palm}. In this paper, we investigate the use of diffusion models for audio-visual pre-training along with various strategies (e.g., curriculumn-based masking) to improve pre-training efficiency. \cref{fig:diffmavil} shows the overview of our model.

\textbf{Contributions.} We make the following contributions: 1) We show that diffusion-based masked audio-video pre-training can facilitate rich audio-video representation learning in downstream audio classification tasks while being more amenable to efficiency optimization strategies (\cref{section:main_results} and \cref{section:ablations}). 2) We show that pre-training computational efficiency can be improved without compromising performance by using cross-attention instead of self-attention (\cref{section:ablations}). We study a masking ratio curriculum along with a dynamic batch size that reduces pre-training FLOPS by 32\% and wall-clock pre-training time by 18\% (\cref{section:ablations}) while maintaining accuracy.

\section{Related Work}
The MAE framework was introduced in the context of image representation learning \cite{ImageMAE} and has been extended to multiple modalities, including audio \cite{AudioMAE}, and video \cite{SpatiotemporalMAE, VideoMAE, VideoMAEv2}. MAE models are ViT-based encoder-decoder architectures that aim to learn feature representations by randomly masking a large fraction of patches and attempting to reconstruct masked patches from visible latents. MAViL \cite{MAViL} is a two-stage self-supervised audio-video representation learning framework built upon the MAE framework. It aims to learn a joint audio-video latent space by leveraging contrastive learning and knowledge distillation techniques as an extension of MAEs. In the first stage, MAViL's objective is to minimize audio and video reconstruction errors as in conventional MAEs; however, this first stage jointly facilitates alignment within and across modalities by minimizing the InfoNCE contrastive loss \cite{InfoNCE} under different ``views'' of the same instance for within-modal alignment, and by minimizing the InfoNCE loss under different modality embeddings derived from the same instance. MAViL allows for an optional second stage, in which knowledge distillation is used to train a student MAViL model on the outputs of a teacher MAViL model trained during the first stage. To enable fair comparisons, and to avoid doubling compute requirements, we train all methods without distillation, i.e., we pre-train only the first stage.

DiffMAE \cite{DiffusionMAE} introduced diffusion into the MAE framework. Rather than append \mask tokens to the visible patch embeddings output by the MAE encoder, DiffMAE diffuses the masked patches and appends these to the visible patch embeddings which are then fed through the MAE decoder. In the next section, we introduce our DiffMAViL framework, which integrates diffusion into MAViL and incorporates several strategies to improve efficiency.

\section{DiffMAViL}\label{section:diffmavil}

\paragraph{Model.} To encourage our model to learn representations that capture high frequency features, and motivated by DiffMAE \cite{DiffusionMAE}, we augment the MAViL audio and video branches with diffusion \cite{ddpm}. Our approach is outlined in \cref{fig:diffmavil}. Contrary to MAViL, which appends \mask tokens to the multi-modal audio and video representations output by the fusion encoder, we instead diffuse the masked patches, project them, and then append these diffused patches to the multi-modal embeddings, which are then fed into their corresponding modality decoders. For a masked (audio or video) patch $x_0^m$, the diffused patch is given by $x_t^m \sim \gN(\sqrt{\bar{\alpha}_t}x_0^m, (1-\bar{\alpha}_t)\rmI)$ where $t \sim \textrm{Unif}([1, 2, \ldots, T])$ is a randomly-sampled timestep and $\bar{\alpha}_t = \prod_{i=1}^t (1-\beta_t)$ where $\beta_{1:T}$ is the variance schedule. We refer to our MAViL with diffusion model as \textit{DiffMAViL} and provide additional details in \cref{sec:appendix_background_diffmavil}.

\paragraph{Training efficiency.} To improve the training efficiency of DiffMAViL, we study following methods:

\begin{itemize}[leftmargin=*]
    \item \emph{Cross-Attention.} We begin by replacing the self-attention modules in our video branch's decoder with cross-attention modules \cite{vaswani2017attention}. In cross-attention,  masked patch embeddings only attend to visible patch embeddings. Masked patch embeddings constitute the ``query’’ sequence, while the visible patch embeddings comprise the ``key-value’’ sequence. Due to transformers’ quadratic complexity in the sequence length, cross-attention is more efficient than self-attention which operates on the concatenated sequence of masked and visible patch embeddings.  Our decoder is similar to the ``cross'' decoder presented in \cite{DiffusionMAE}, however, our cross-attention modules attend only to the visible latents of the final encoder block, rather than to all of them. For the audio decoder, we use the Swin-Transformer local attention \cite{swintrans} as this was shown to perform favorably in \cite{AudioMAE}.
    \item \emph{Masking Ratio Curriculum.} Curriculum learning \cite{bengio2009curriculum} aims to organize training samples in a way that facilitates learning. This notion has inspired several progressive learning methods \cite{efficientnetv2, MSCVBSWC} that progressively increase the resolution of training samples throughout training. Inspired by this, we propose a dynamic masking ratio that progressively  decays over the course of training. In MAViL, a fixed masking ratio, $\rho \in (0,1)$, is used throughout training. As the transformer blocks for both the audio and video encoders in DiffMAViL operate only on visible patches, we can improve efficiency by processing fewer visible patches. The number of visible patches is a fraction, $1 - \rho$, of the total number of patches. Hence, by using a larger value of $\rho$, we mask out a greater number of patches and consequently process fewer visible patches. We therefore propose having a dynamic masking ratio that begins at $\rho_1 \in (0,1)$ and ends at $\rho_2 \in (0,1)$ following a schedule. We consider a simple linear masking ratio schedule that varies from $\rho_1$ at the start of training to $\rho_2$ at the end of training. 
    \item \emph{Adaptive Batch Size.} In vision tasks, training with a lower sample resolution naturally entails the utilization of fewer computational resources, which may lead to underutilization of accelerators. In an effort to combat this, several works \cite{MSCVBSWC, CVNets, MobileViT} have used an adaptive batch size, where larger batch sizes are used when training at a lower resolution, and smaller batch sizes are used when training at a higher resolution, resulting in faster training. We extend these methods by making the batch size adaptive to the masking ratio. For a base batch size, $B_0$ (i.e., the batch size that will be used for the masking ratio $\min(\rho_1, \rho_2)$), the batch size at epoch $e$ is given by $B_e = \frac{1 - \min(\rho_1, \rho_2)}{1 - \rho_e}B_0$, where $\rho_e$ is the masking ratio at epoch $e$ as determined by the masking ratio schedule.
\end{itemize}

\section{Experiments}
To pre-train our models, we use the union of the ``balanced'' and ``unbalanced'' splits of the  AudioSet \cite{audioset} dataset, denoted ``AS-2M.'' We note that we were only able to acquire 85\% of the total AudioSet dataset, as many videos are no longer available on YouTube. We pre-train all baselines on this dataset for fair comparison. We focus on fine-tuning on only the audio modality, i.e., we fine-tune only the audio encoder branch of our DiffMAViL and MAViL models. We fine-tune on the ``balanced'' AudioSet split (denoted ``AS-20K'') and report the mean average precision (mAP). Additionally, we fine-tune on VGGSound \cite{vggsound}, Environmental Sound Classification (ESC-50) \cite{esc50}, and Speech Commands v2 (SPC-v2) \cite{SPC} where we use the split considered in \cite{convmixerSPC}. We report the Top-1 (\%) accuracy for VGGSound, ESC-50, and SPC-v2. For ESC-50, we report the mean accuracy under standard five-fold cross validation. For each experiment, we report the mean and standard deviation of three independent seeds. Additional training details are provided in \cref{section:appendix_training_details}.

\begin{table*}[h!]
    \centering
    \resizebox{\columnwidth}{!}{
    \begin{tabular}{c|cccc|cc}
        \toprule[1.5pt]
        \textbf{Model} & \makecell{\textbf{AS-20K} \\ \small{(mAP $\uparrow$)}} & \makecell{\textbf{VGGSound} \\ \small{(Top-1 $\uparrow$)}}  & \makecell{\textbf{ESC-50} \\ \small{(Top-1 $\uparrow$)}}   & \makecell{\textbf{SPC-v2} \\ \small{(Top-1 $\uparrow$)}}  & \textbf{FLOPS} & \makecell{\textbf{Avg. Epoch} \\ \textbf{Time}} \\
        \midrule[1.25pt]
        MAViL$^*$ \cite{MAViL} & $35.9 \pm 0.10$ & $57.2 \pm 0.12$ & $93.7 \pm 0.13$ & $98.0 \pm 0.08$ & $1\times$ & $1\times$ \\ 
        DiffMAViL (ours) & $35.8 \pm 0.16$ & $57.0 \pm 0.17$ & $93.1 \pm 0.18$ & $97.7 \pm 0.04$ & $\mathbf{0.68}\times$ & $\mathbf{0.82}\times$ \\ 
        \bottomrule[1.5pt]
    \end{tabular}
    }
\caption{\textbf{DiffMAViL improves training efficiency while maintaining accuracy.} Our DiffMAViL model integrates diffusion into the MAViL \cite{MAViL} framework along with a cross-attention video decoder, linear masking ratio schedule, and a dynamic batch size to improve efficiency. $^*$We present results for our own MAViL implementation as the public release is not available at the time of writing.}
\label{table:mavil_vs_diffmavil}
\end{table*}

\begin{table*}[h!]
    \centering
    \resizebox{\columnwidth}{!}{
    \begin{tabular}{cccc|cccc|cc}
        \toprule[1.5pt]
        \textbf{Row \#} &  \makecell{\textbf{Video} \\ \textbf{attention}} & \makecell{\textbf{Masking} \\ \textbf{ratio}} &  \makecell{\textbf{Adaptive} \\ \textbf{batch size}} & \makecell{\textbf{AS-20K} \\ \small{(mAP $\uparrow$)}} & \makecell{\textbf{VGGSound} \\ \small{(Top-1 $\uparrow$)}}  & \makecell{\textbf{ESC-50} \\ \small{(Top-1 $\uparrow$)}}   & \makecell{\textbf{SPC-v2} \\ \small{(Top-1 $\uparrow$)}}  & \textbf{FLOPS} & \makecell{\textbf{Avg. epoch} \\ \textbf{time}} \\
        \midrule[1.25pt]
        R1 & Self & Fixed & \xmark & $36.0 \pm 0.08$ & $57.5 \pm 0.09$ & $94.7 \pm 0.10$ & $97.9 \pm 0.04$ & $1\times$ &  $1\times$ \\ 
        R2 & \textcolor{Magenta}{Cross} & Fixed & \xmark & $36.3 \pm 0.09$ & $57.4 \pm 0.03$ & $94.2 \pm 0.20$ & $97.9 \pm 0.08$ & \textcolor{Magenta}{$0.81\times$} & $0.96 \times$\\ 
        R3 & Cross & \textcolor{NavyBlue}{Linear} & \xmark & $36.0 \pm 0.07$ & $57.3 \pm 0.19$ & $93.3 \pm 0.10$ & $97.6 \pm 0.05$ & \textcolor{NavyBlue}{$0.68\times$} &  $0.96 \times$ \\ 
        R4 & Cross & Linear & \textcolor{ForestGreen}{\cmark} & $35.8 \pm 0.16$ & $57.0 \pm 0.17$ & $93.1 \pm 0.18$ & $97.7 \pm 0.04$ & $0.68\times$ & \textcolor{ForestGreen}{$0.82\times$} \\ 
        \bottomrule[1.5pt]
    \end{tabular}
    }
\caption{\textbf{DiffMAViL ablations.} Compared to our baseline DiffMAViL model (R1), replacing the video decoder's self-attention modules with cross-attention reduces pre-training FLOPS by 19\% (R2). Replacing the fixed masking ratio of 0.8 with a linear schedule that decays from 0.9 to 0.8 reduces FLOPS by 32\% (R3). Adding an adaptive batch size reduces pre-training wall-clock time by 18\% (R4).}
\label{table:diffmavil_ablations}
\end{table*}

\subsection{Main Results}\label{section:main_results}
In \cref{table:mavil_vs_diffmavil}, we compare the performance of MAViL against our DiffMAViL model. We observe that the use of diffusion, coupled with our efficiency strategies outlined in \cref{section:diffmavil}, reduces pre-training FLOPS and wall-clock time without incurring a significant loss in performance. In \cref{section:appendix_audio_mae_and_diffusion}, we show that augmenting AudioMAE \cite{AudioMAE} with diffusion improves downstream performance, suggesting that diffusion may aid in the learning of richer audio representations.

\subsection{Ablations}\label{section:ablations}
\paragraph{Replacing Self-Attention with Cross-Attention.} To reduce pre-training compute, we replace the self-attention modules in the video branch's decoder with cross-attention \cite{vaswani2017attention}. In row R2 of \cref{table:diffmavil_ablations}, we observe that the use of cross-attention reduces pre-training FLOPS by 19\% while preserving accuracy across multiple datasets. In \cref{section:appendix_mavil_vs_diffmavil_cross}, we show that DiffMAViL is more amenable to cross-attention than MAViL; we therefore focus our efforts on improving the efficiency of DiffMAViL.

\paragraph{Curriculum-Based Masking.} We further improve training efficiency by augmenting DiffMAViL with a curriculum for the masking ratio. In row R3 of \cref{table:diffmavil_ablations}, we show that a linear schedule that decays the masking ratio from 0.9 to 0.8 throughout training reduces pre-training FLOPS by 32\%. In \cref{section:appendix_flop_analysis}, we analyze the FLOPS reduction within each encoder and decoder module.

\paragraph{Adaptive Batch Size.} 
While pre-training with a dynamic masking ratio reduces pre-training FLOPS, it does not have a significant decrease in the wall-clock training time as it leads to under-utilization of computational resources. To offset this, we augment DiffMAViL with an adaptive batch size in service of maintaining constant compute at each iteration (\cref{section:diffmavil} for details). The dynamic balance between masking ratio and batch size allows us to utilize hardware more efficiently and maintain similar FLOPs to R3 in \cref{table:diffmavil_ablations}. Consequently, this reduces the number of optimization steps required per epoch, resulting in an 18\% reduction in wall-clock pre-training time while maintaining accuracy.

\section{Conclusion}
We study DiffMAViL, an audio-video pre-training framework with diffusion. Our results shows that the integration of diffusion techniques into MAViL, along with the implementation of diverse training efficiency strategies, such as a masking ratio curriculum and adaptive batch size, leads to a significant reduction of 32\% in pre-training FLOPS and an 18\% decrease in pre-training wall clock time. 

\bibliography{refs}

\begin{thebibliography}{41}
\providecommand{\natexlab}[1]{#1}
\providecommand{\url}[1]{\texttt{#1}}
\expandafter\ifx\csname urlstyle\endcsname\relax
  \providecommand{\doi}[1]{doi: #1}\else
  \providecommand{\doi}{doi: \begingroup \urlstyle{rm}\Url}\fi

\bibitem[Huang et~al.(2022{\natexlab{a}})Huang, Sharma, Xu, Ryali, Fan, Li, Li, Ghosh, Malik, and Feichtenhofer]{MAViL}
Po-Yao Huang, Vasu Sharma, Hu~Xu, Chaitanya Ryali, Haoqi Fan, Yanghao Li, Shang-Wen Li, Gargi Ghosh, Jitendra Malik, and Christoph Feichtenhofer.
\newblock Mavil: Masked audio-video learners.
\newblock \emph{arXiv preprint arXiv:2212.08071}, 2022{\natexlab{a}}.

\bibitem[Gong et~al.(2022)Gong, Rouditchenko, Liu, Harwath, Karlinsky, Kuehne, and Glass]{CavMAE}
Yuan Gong, Andrew Rouditchenko, Alexander~H Liu, David Harwath, Leonid Karlinsky, Hilde Kuehne, and James~R Glass.
\newblock Contrastive audio-visual masked autoencoder.
\newblock In \emph{The Eleventh International Conference on Learning Representations}, 2022.

\bibitem[Aytar et~al.(2016)Aytar, Vondrick, and Torralba]{soundnet}
Yusuf Aytar, Carl Vondrick, and Antonio Torralba.
\newblock Soundnet: Learning sound representations from unlabeled video.
\newblock \emph{Advances in neural information processing systems}, 29, 2016.

\bibitem[Korbar et~al.(2018)Korbar, Tran, and Torresani]{korbar2018cooperative}
Bruno Korbar, Du~Tran, and Lorenzo Torresani.
\newblock Cooperative learning of audio and video models from self-supervised synchronization.
\newblock \emph{Advances in Neural Information Processing Systems}, 31, 2018.

\bibitem[Arandjelovic and Zisserman(2017)]{looklistenlearn}
Relja Arandjelovic and Andrew Zisserman.
\newblock Look, listen and learn.
\newblock In \emph{Proceedings of the IEEE international conference on computer vision}, pages 609--617, 2017.

\bibitem[Doersch et~al.(2015)Doersch, Gupta, and Efros]{doersch2015unsupervised}
Carl Doersch, Abhinav Gupta, and Alexei~A Efros.
\newblock Unsupervised visual representation learning by context prediction.
\newblock In \emph{Proceedings of the IEEE international conference on computer vision}, pages 1422--1430, 2015.

\bibitem[Wang and Gupta(2015)]{wang2015unsupervised}
Xiaolong Wang and Abhinav Gupta.
\newblock Unsupervised learning of visual representations using videos.
\newblock In \emph{Proceedings of the IEEE international conference on computer vision}, pages 2794--2802, 2015.

\bibitem[Noroozi and Favaro(2016)]{noroozi2016unsupervised}
Mehdi Noroozi and Paolo Favaro.
\newblock Unsupervised learning of visual representations by solving jigsaw puzzles.
\newblock In \emph{European conference on computer vision}, pages 69--84. Springer, 2016.

\bibitem[Gidaris et~al.(2018)Gidaris, Singh, and Komodakis]{gidaris2018unsupervised}
Spyros Gidaris, Praveer Singh, and Nikos Komodakis.
\newblock Unsupervised representation learning by predicting image rotations.
\newblock \emph{arXiv preprint arXiv:1803.07728}, 2018.

\bibitem[Vincent et~al.(2008)Vincent, Larochelle, Bengio, and Manzagol]{denoisingautoencoder}
Pascal Vincent, Hugo Larochelle, Yoshua Bengio, and Pierre-Antoine Manzagol.
\newblock Extracting and composing robust features with denoising autoencoders.
\newblock In \emph{Proceedings of the 25th international conference on Machine learning}, pages 1096--1103, 2008.

\bibitem[He et~al.(2022)He, Chen, Xie, Li, Doll{\'a}r, and Girshick]{ImageMAE}
Kaiming He, Xinlei Chen, Saining Xie, Yanghao Li, Piotr Doll{\'a}r, and Ross Girshick.
\newblock Masked autoencoders are scalable vision learners.
\newblock In \emph{Proceedings of the IEEE/CVF conference on computer vision and pattern recognition}, pages 16000--16009, 2022.

\bibitem[Feichtenhofer et~al.(2022)Feichtenhofer, Li, He, et~al.]{SpatiotemporalMAE}
Christoph Feichtenhofer, Yanghao Li, Kaiming He, et~al.
\newblock Masked autoencoders as spatiotemporal learners.
\newblock \emph{Advances in neural information processing systems}, 35:\penalty0 35946--35958, 2022.

\bibitem[Tong et~al.(2022)Tong, Song, Wang, and Wang]{VideoMAE}
Zhan Tong, Yibing Song, Jue Wang, and Limin Wang.
\newblock Videomae: Masked autoencoders are data-efficient learners for self-supervised video pre-training.
\newblock \emph{Advances in neural information processing systems}, 35:\penalty0 10078--10093, 2022.

\bibitem[Wang et~al.(2023)Wang, Huang, Zhao, Tong, He, Wang, Wang, and Qiao]{VideoMAEv2}
Limin Wang, Bingkun Huang, Zhiyu Zhao, Zhan Tong, Yinan He, Yi~Wang, Yali Wang, and Yu~Qiao.
\newblock Videomae v2: Scaling video masked autoencoders with dual masking.
\newblock In \emph{Proceedings of the IEEE/CVF Conference on Computer Vision and Pattern Recognition}, pages 14549--14560, 2023.

\bibitem[Huang et~al.(2022{\natexlab{b}})Huang, Xu, Li, Baevski, Auli, Galuba, Metze, and Feichtenhofer]{AudioMAE}
Po-Yao Huang, Hu~Xu, Juncheng Li, Alexei Baevski, Michael Auli, Wojciech Galuba, Florian Metze, and Christoph Feichtenhofer.
\newblock Masked autoencoders that listen.
\newblock \emph{Advances in Neural Information Processing Systems}, 35:\penalty0 28708--28720, 2022{\natexlab{b}}.

\bibitem[Wei et~al.(2023)Wei, Mangalam, Huang, Li, Fan, Xu, Wang, Xie, Yuille, and Feichtenhofer]{DiffusionMAE}
Chen Wei, Karttikeya Mangalam, Po-Yao Huang, Yanghao Li, Haoqi Fan, Hu~Xu, Huiyu Wang, Cihang Xie, Alan Yuille, and Christoph Feichtenhofer.
\newblock Diffusion models as masked autoencoders.
\newblock \emph{arXiv preprint arXiv:2304.03283}, 2023.

\bibitem[Rombach et~al.(2022)Rombach, Blattmann, Lorenz, Esser, and Ommer]{rombach2022high}
Robin Rombach, Andreas Blattmann, Dominik Lorenz, Patrick Esser, and Bj{\"o}rn Ommer.
\newblock High-resolution image synthesis with latent diffusion models.
\newblock In \emph{Proceedings of the IEEE/CVF conference on computer vision and pattern recognition}, pages 10684--10695, 2022.

\bibitem[Dhariwal and Nichol(2021)]{dhariwal2021diffusion}
Prafulla Dhariwal and Alexander Nichol.
\newblock Diffusion models beat gans on image synthesis.
\newblock \emph{Advances in neural information processing systems}, 34:\penalty0 8780--8794, 2021.

\bibitem[Chowdhery et~al.(2022)Chowdhery, Narang, Devlin, Bosma, Mishra, Roberts, Barham, Chung, Sutton, Gehrmann, et~al.]{chowdhery2022palm}
Aakanksha Chowdhery, Sharan Narang, Jacob Devlin, Maarten Bosma, Gaurav Mishra, Adam Roberts, Paul Barham, Hyung~Won Chung, Charles Sutton, Sebastian Gehrmann, et~al.
\newblock Palm: Scaling language modeling with pathways.
\newblock \emph{arXiv preprint arXiv:2204.02311}, 2022.

\bibitem[Oord et~al.(2018)Oord, Li, and Vinyals]{InfoNCE}
Aaron van~den Oord, Yazhe Li, and Oriol Vinyals.
\newblock Representation learning with contrastive predictive coding.
\newblock \emph{arXiv preprint arXiv:1807.03748}, 2018.

\bibitem[Ho et~al.(2020)Ho, Jain, and Abbeel]{ddpm}
Jonathan Ho, Ajay Jain, and Pieter Abbeel.
\newblock Denoising diffusion probabilistic models.
\newblock \emph{Advances in neural information processing systems}, 33:\penalty0 6840--6851, 2020.

\bibitem[Vaswani et~al.(2017)Vaswani, Shazeer, Parmar, Uszkoreit, Jones, Gomez, Kaiser, and Polosukhin]{vaswani2017attention}
Ashish Vaswani, Noam Shazeer, Niki Parmar, Jakob Uszkoreit, Llion Jones, Aidan~N Gomez, {\L}ukasz Kaiser, and Illia Polosukhin.
\newblock Attention is all you need.
\newblock \emph{Advances in neural information processing systems}, 30, 2017.

\bibitem[Liu et~al.(2021)Liu, Lin, Cao, Hu, Wei, Zhang, Lin, and Guo]{swintrans}
Ze~Liu, Yutong Lin, Yue Cao, Han Hu, Yixuan Wei, Zheng Zhang, Stephen Lin, and Baining Guo.
\newblock Swin transformer: Hierarchical vision transformer using shifted windows.
\newblock In \emph{Proceedings of the IEEE/CVF international conference on computer vision}, pages 10012--10022, 2021.

\bibitem[Bengio et~al.(2009)Bengio, Louradour, Collobert, and Weston]{bengio2009curriculum}
Yoshua Bengio, J{\'e}r{\^o}me Louradour, Ronan Collobert, and Jason Weston.
\newblock Curriculum learning.
\newblock In \emph{Proceedings of the 26th annual international conference on machine learning}, pages 41--48, 2009.

\bibitem[Tan and Le(2021)]{efficientnetv2}
Mingxing Tan and Quoc Le.
\newblock Efficientnetv2: Smaller models and faster training.
\newblock In \emph{International conference on machine learning}, pages 10096--10106. PMLR, 2021.

\bibitem[Nunez et~al.(2023)Nunez, Merth, Prabhu, Farajtabar, Rastegari, Mehta, and Horton]{MSCVBSWC}
Elvis Nunez, Thomas Merth, Anish Prabhu, Mehrdad Farajtabar, Mohammad Rastegari, Sachin Mehta, and Maxwell Horton.
\newblock On the efficacy of multi-scale data samplers for vision applications.
\newblock \emph{arXiv preprint arXiv:2309.04502}, 2023.

\bibitem[Mehta et~al.(2022)Mehta, Abdolhosseini, and Rastegari]{CVNets}
Sachin Mehta, Farzad Abdolhosseini, and Mohammad Rastegari.
\newblock Cvnets: High performance library for computer vision.
\newblock In \emph{Proceedings of the 30th ACM International Conference on Multimedia}, pages 7327--7330, 2022.

\bibitem[Mehta and Rastegari(2021)]{MobileViT}
Sachin Mehta and Mohammad Rastegari.
\newblock Mobilevit: Light-weight, general-purpose, and mobile-friendly vision transformer.
\newblock In \emph{International Conference on Learning Representations}, 2021.

\bibitem[Gemmeke et~al.(2017)Gemmeke, Ellis, Freedman, Jansen, Lawrence, Moore, Plakal, and Ritter]{audioset}
Jort~F Gemmeke, Daniel~PW Ellis, Dylan Freedman, Aren Jansen, Wade Lawrence, R~Channing Moore, Manoj Plakal, and Marvin Ritter.
\newblock Audio set: An ontology and human-labeled dataset for audio events.
\newblock In \emph{2017 IEEE international conference on acoustics, speech and signal processing (ICASSP)}, pages 776--780. IEEE, 2017.

\bibitem[Chen et~al.(2020)Chen, Xie, Vedaldi, and Zisserman]{vggsound}
Honglie Chen, Weidi Xie, Andrea Vedaldi, and Andrew Zisserman.
\newblock Vggsound: A large-scale audio-visual dataset.
\newblock In \emph{ICASSP 2020-2020 IEEE International Conference on Acoustics, Speech and Signal Processing (ICASSP)}, pages 721--725. IEEE, 2020.

\bibitem[Piczak(2015)]{esc50}
Karol~J Piczak.
\newblock Esc: Dataset for environmental sound classification.
\newblock In \emph{Proceedings of the 23rd ACM international conference on Multimedia}, pages 1015--1018, 2015.

\bibitem[Warden(2018)]{SPC}
Pete Warden.
\newblock Speech commands: A dataset for limited-vocabulary speech recognition.
\newblock \emph{arXiv preprint arXiv:1804.03209}, 2018.

\bibitem[Ng et~al.(2022)Ng, Chen, Tian, Fu, and Chng]{convmixerSPC}
Dianwen Ng, Yunqi Chen, Biao Tian, Qiang Fu, and Eng~Siong Chng.
\newblock Convmixer: Feature interactive convolution with curriculum learning for small footprint and noisy far-field keyword spotting.
\newblock \emph{arXiv preprint arXiv:2201.05863}, 2022.

\bibitem[Dosovitskiy et~al.(2020)Dosovitskiy, Beyer, Kolesnikov, Weissenborn, Zhai, Unterthiner, Dehghani, Minderer, Heigold, Gelly, et~al.]{vit}
Alexey Dosovitskiy, Lucas Beyer, Alexander Kolesnikov, Dirk Weissenborn, Xiaohua Zhai, Thomas Unterthiner, Mostafa Dehghani, Matthias Minderer, Georg Heigold, Sylvain Gelly, et~al.
\newblock An image is worth 16x16 words: Transformers for image recognition at scale.
\newblock \emph{arXiv preprint arXiv:2010.11929}, 2020.

\bibitem[Loshchilov and Hutter(2017)]{adamw}
Ilya Loshchilov and Frank Hutter.
\newblock Decoupled weight decay regularization.
\newblock In \emph{International Conference on Learning Representations}, 2017.

\bibitem[Loshchilov and Hutter(2016)]{loshchilov2016sgdr}
Ilya Loshchilov and Frank Hutter.
\newblock Sgdr: Stochastic gradient descent with warm restarts.
\newblock \emph{arXiv preprint arXiv:1608.03983}, 2016.

\bibitem[Bao et~al.(2021)Bao, Dong, Piao, and Wei]{beit}
Hangbo Bao, Li~Dong, Songhao Piao, and Furu Wei.
\newblock Beit: Bert pre-training of image transformers.
\newblock \emph{arXiv preprint arXiv:2106.08254}, 2021.

\bibitem[Park et~al.(2019)Park, Chan, Zhang, Chiu, Zoph, Cubuk, and Le]{specaug}
Daniel~S Park, William Chan, Yu~Zhang, Chung-Cheng Chiu, Barret Zoph, Ekin~D Cubuk, and Quoc~V Le.
\newblock Specaugment: A simple data augmentation method for automatic speech recognition.
\newblock \emph{arXiv preprint arXiv:1904.08779}, 2019.

\bibitem[Huang et~al.(2016)Huang, Sun, Liu, Sedra, and Weinberger]{stochasticdropout}
Gao Huang, Yu~Sun, Zhuang Liu, Daniel Sedra, and Kilian~Q Weinberger.
\newblock Deep networks with stochastic depth.
\newblock In \emph{Computer Vision--ECCV 2016: 14th European Conference, Amsterdam, The Netherlands, October 11--14, 2016, Proceedings, Part IV 14}, pages 646--661. Springer, 2016.

\bibitem[Zhang et~al.(2017)Zhang, Cisse, Dauphin, and Lopez-Paz]{mixup}
Hongyi Zhang, Moustapha Cisse, Yann~N Dauphin, and David Lopez-Paz.
\newblock mixup: Beyond empirical risk minimization.
\newblock \emph{arXiv preprint arXiv:1710.09412}, 2017.

\bibitem[Yun et~al.(2019)Yun, Han, Oh, Chun, Choe, and Yoo]{cutmix}
Sangdoo Yun, Dongyoon Han, Seong~Joon Oh, Sanghyuk Chun, Junsuk Choe, and Youngjoon Yoo.
\newblock Cutmix: Regularization strategy to train strong classifiers with localizable features.
\newblock In \emph{Proceedings of the IEEE/CVF international conference on computer vision}, pages 6023--6032, 2019.

\end{thebibliography}

\clearpage
\appendix

\section{MAViL and DiffMAViL with Cross-Attention}\label{section:appendix_mavil_vs_diffmavil_cross}
In this section, we replace self-attention modules with cross-attention in  both MAViL and DiffMAViL. In \cref{table:mavil_vs_diffmavil_cross}, we observe that DiffMAViL with cross-attention tends to have a stronger performance on downstream audio classification tasks compared to MAViL with cross-attention. 

\begin{table*}[h!]
    \centering
    \resizebox{0.7\columnwidth}{!}{
    \begin{tabular}{cc|ccc}
        \toprule[1.5pt]
        \textbf{Model} & \makecell{\textbf{Video} \\ \textbf{Attention}} & \makecell{\textbf{AS-20K} \\ \small{(mAP $\uparrow$)}} & \makecell{\textbf{ESC-50} \\ \small{(Top-1 $\uparrow$)}}   & \makecell{\textbf{SPC-v2} \\ \small{(Top-1 $\uparrow$)}} \\
        \midrule[1.25pt]
        MAViL & cross & $36.1 \pm 0.04$ & $93.6 \pm 0.25$ & $98.0 \pm 0.12$ \\ 
        DiffMAViL (ours) & cross & $36.3 \pm 0.09$ & $94.2 \pm 0.20$ & $97.9 \pm 0.08$ \\ 
        \bottomrule[1.5pt]
    \end{tabular}
    }
\caption{\textbf{DiffMAViL Is More Amenable to Cross-Attention.} Replacing self-attention in DiffMAViL's video decoder with cross-attention has a more positive effect on downstream performance compared to MAViL with cross-attention. Efficiency metrics are measured relative to the standard MAViL model in \cref{table:mavil_vs_diffmavil}.}
\label{table:mavil_vs_diffmavil_cross}
\end{table*}

\section{MAViL and DiffMAViL Background}\label{sec:appendix_background}
In this section, we provide additional details regarding MAViL's first training stage, as well as the diffusion process of our DiffMAViL model and our use of cross-attention.

\subsection{MAViL}\label{sec:appendix_background_mavil}
Let $(a, v)$ be an audio-video instance pair where $a$ is an audio spectrogram and $v$ is a tensor of RGB video frames. $a$ and $v$ are first patchified and tokenized, producing $\rva=[a_1, \ldots,a_M]$ audio tokens and $\rvv=[v_1, \ldots, v_N]$ video tokens where $a_i, v_j \in \R^d$. In the encoding step, a fraction, $\rho \in (0,1)$, of the audio and video tokens are then randomly masked, yielding $\rva'$ and $\rvv'$ containing $\lfloor (1-\rho) M \rceil$ and $\lfloor (1-\rho) N \rceil$ visible tokens, respectively, where $\lfloor \cdot \rceil$ denotes rounding to the nearest integer. These visible tokens are then embedded by audio and video ViT-based encoders, $f_a$, and $f_v$, producing the uni-modal audio and video representations $\rva_{um} \triangleq f_a(\rva')$ and $\rvv_{um} \triangleq f_v(\rvv')$. The uni-modal representations are then concatenated, forming $(\rva_{um},\rvv_{um})$, and passed through a ViT-based fusion encoder, $g_{av}$, producing multi-modal representations $(\rva_{mm},\rvv_{mm}) \triangleq g_{av}(\rva_{um},\rvv_{um})$. In the decoding step, the $\rva_{mm}$ and $\rvv_{mm}$ are first projected onto the decoder space. Then, a learnable \mask token is appended to each of the multi-modal representations for each of the masked patches in the encoding step, yielding  $\tilde{\rva}_{mm}$ and $\tilde{\rvv}_{mm}$. These are then passed through ViT-based decoders for each modality, denoted $f^{-1}_a$ and $f^{-1}_v$, followed by a linear projection head $l_a$ and $l_v$. Therefore, the reconstructions of (patchified) $a$ and $v$ are given by $\hat{\rva} \triangleq l_a(f^{-1}_a(\tilde{\rva}_{mm}))$ and $\hat{\rvv} \triangleq l_v(f^{-1}_v(\tilde{\rvv}_{mm}))$. Letting $\rva_i^{\raw}$, $i=1, \ldots, M$, and $\rvv_j^{\raw}$, $j=1, \ldots, N$ denote the patches of the original audio and video inputs, the mean-squared error (MSE) loss is given by $\Ls^{MSE} = \frac{1}{M}\sum_{i=1}^M (\hat{\rva}_i - \rva_i)^2 + \frac{1}{N}\sum_{j=1}^N (\hat{\rvv}_j - \rvv_j)^2$.

In addition to minimizing the MSE loss, the first stage of MAViL also considers two contrastive losses. The first, the ``inter-modal'' loss, facilitates alignment across modalities by first averaging the audio and video uni-modal representations, $\rva_{emb} \triangleq \textrm{Avg}(\rva_{um})$, $\rvv_{emb} \triangleq \textrm{Avg}(\rvv_{um})$, where $\textrm{Avg}(\cdot)$ denotes averaging along the sequence length. These instance-level representations are then fed through the InfoNCE loss, where video and audio clips from the same video constitute positive pairs while all other pairs are negatives. The second loss, the ``intra-modal'' loss, promotes alignment within each modality. By applying a second random masking to the input audio and video clips, a second ``view'' of each modality can be obtained, $\bar{\rva}_{emb}$ and $\bar{\rvv}_{emb}$, which are then also fed through the InfoNCE loss. In this case, the two views from the same instance are considered a positive pair and the negative pairs consist of the views from all other instances of the same modality. MAViL's first stage objective function is therefore a linear combination of the MSE loss and the two contrastive losses; hence, this procedure consists of four forward passes through the encoders (one pass for each view through its respective modality's encoder).

\subsection{DiffMAViL}\label{sec:appendix_background_diffmavil}
In MAViL's audio and video decoders, learnable \mask tokens are used to represent masked spectogram/RGB frame patches. In our DiffMAViL model, we replace the learnable \mask tokens with diffused patches. Let $x_0^m$ represent a masked audio or video frame patch where $m$ denotes a masked patch and the subscript denotes the diffusion time step. At each training iteration, we sample $t \sim \textrm{Unif}(\{1, 2, \ldots, T\})$, and diffuse $x_0^m$ according to noise level $t$ to obtain $x_t^m = \sqrt{1-\bar{\alpha}_t}\epsilon + \sqrt{\bar{\alpha}_t} x_0^m$ where $\epsilon \sim \gN(0, \mathbf{I})$ is a standard normal sample with the same dimension as $x_0^m$. The multi-modal embeddings output by the fusion encoder, along with $x_t^m$, are then projected onto the decoder's embedding space, and after restoring the original patch ordering, are fed through the corresponding decoder. The decoders are therefore tasked with reconstructing the original input from the visible patch emebeddings and diffused masked patches in a single step.

As in \cite{DiffusionMAE}, when training with diffusion we use the ``simple'' objective function proposed in \cite{ddpm}. Namely, the objective is to minimize the reconstruction error between the masked input $x_0^m$, and the decoder's reconstruction given $x_t^m$ and the visible latents. In other words, this reduces to the reconstruction MSE used by MAViL. The objective function optimized by DiffMAViL is therefore the same as MAViL.

We also explored the use of cross-attention instead of self-attention in the video decoder in order to improve efficiency. Typically, self-attention is applied to the concatenated sequence of masked and unmasked patch embeddings. When using cross-attention, unmasked patch embeddings attend to masked patch embeddings. We do no not apply cross-attention to the audio decoder as local attention was shown to have strong performance for audio \cite{AudioMAE}; however, future work can explore the use of cross-attention in the audio decoder as well.

\section{Training Details}\label{section:appendix_training_details}
Our audio encoder-decoder architecture follows that of AudioMAE \cite{AudioMAE}, while our video encoder-decoder architecture follows that of SpatiotemporalMAE \cite{SpatiotemporalMAE}. Namely, our audio and video encoders are both ViT-B models \cite{vit}. Both decoders have 8 transformer blocks, 16 attention heads, and an embedding dimension of 512. The audio decoder uses local attention Swin-Transformer \cite{swintrans} blocks. Both encoders and decoders use sinusoidal positional embeddings, and the video encoder and decoder use separable temporal and spatial positional embeddings. The fusion encoder consists of a two-layer Transformer. As a masking ratio of 0.8 was shown to perform well in \cite{MAViL}, we also use a masking ratio of 0.8 as our default. Notably, we pre-train both our DiffMAViL and the standard MAViL models with the same hyperparameters. Moreover, as the the code for MAViL is not publicly available at the time of writing, our results for this model are from our own implementation.

We pre-train with both audio and video modalities, and fine-tune only on audio tasks. To construct audio spectrograms, we use the entirety of the data sample. For AudioSet and VGGSound, this corresponds to 10 second audio clips. ESC-50 and SPC-v2 correspond to 5 and 1 second clips, respectively. We use a 16K sampling rate and 128 Mel-frequency bands with a 25ms Hanning window shifting every 10ms. This yields spectrograms with shapes $1024\times 128$, $1024\times 128$, $512\times 128$, and $128\times 128$ for AudioSet, VGGSound, ESC-50, and SPC-v2, respectively. For video, we sample 4-second clips consisting of 16 frames. We use a spatial patch size of $16\times 16$ for both audio and video, and a temporal patch size of 2.

In \cref{table:training_details}, we provide the hyperparameters used to train DiffMAViL and MAViL (note that we use the same hyperparameters for both models). For diffusion, we use a linear variance schedule, $\beta_t$, with $t \in \{1, 2, \ldots, 1000\}$. $\beta_t$ increases linearly from $10^{-4}$ to 0.02. As was done in \cite{DiffusionMAE}, we exponentiate the variances with hyperparameter $\phi=0.8$ so that the noise variance is $\beta_t^\phi$. This amplifies the noise used at lower diffusion steps $t$. 

We note that we did not use a weighted sampling for neither pre-training nor fine-tuning on any dataset. All of our training was done on NVIDIA A100 GPUs.

\begin{table*}[ht]
    \centering
    \resizebox{\columnwidth}{!}{
    \begin{tabular}{lccccc}
        \toprule[1.5pt]
        \multirow{2}{*}{\textbf{Configuration}} & \textbf{Pre-training} & \multicolumn{4}{c}{\textbf{Fine-tuning}} \\
         \cmidrule[1.25pt]{3-6}
         & \textbf{AS-2M} & \textbf{AS-20k} & \textbf{VGGSound} & \textbf{ESC-50} & \textbf{SPC-v2} \\
        \midrule[1.25pt]
        Optimizer & \multicolumn{5}{c}{ AdamW \cite{adamw}} \\
        Optimizer momentum & \multicolumn{5}{c}{$\beta_1=0.9, \beta_2=0.95$} \\
        Weight Decay & 1e-5 & 1e-4 & 1e-4 & 1e-4 & 1e-4 \\
        Learning rate & 4e-4 & 2.5e-4 & 2e-4 & 2.5e-4 & 1e-3 \\
        Learning rate schedule & \multicolumn{5}{c}{ Cosine decay \cite{loshchilov2016sgdr}} \\
        Layer-wise learning rate decay \cite{beit} & None & 0.75 & 0.75 & 0.75 & 0.75 \\
        Minimum learning rate & 1e-6 & 1e-6 & 1e-6 & 1e-6 & 1e-6 \\
        Warm-up epochs & 8 & 4 & 1 & 4 & 4 \\
        Epochs & 60 & 60 & 60 & 100 & 60 \\
        Batch size$^*$ & 2048 & 64 & 256 & 64 & 256 \\
        GPUs & 256 & 1 & 4 & 1 & 1\\
        Augmentation$^\dag$ & R & R & R+N & R & R \\
        SpecAug \cite{specaug} (time/freq) & None & 192/48 & 192/48 & 96/24 & 48/48 \\
        Stochastic dropout \cite{stochasticdropout} & 0 & 0.1 & 0.1 & 0.1 & 0.1 \\
        Mixup \cite{mixup} & None & 0.5 & 0.5 & 0 & 0 \\
        Cutmix \cite{cutmix} & None & 1.0 & 1.0 & 0 & 0\\
        Multilabel & - & True & False & False & False \\
        Loss function$^\ddag$ & MSE+Contrastive & BCE & BCE & CE & CE\\
        Dataset mean & -4.268 & -4.268 & -5.189 & -6.627 & -6.702 \\
        Dataset std & 4.569 & 4.569 & 3.260 & 5.359 & 5.448 \\
        \bottomrule[1.5pt]
    \end{tabular}
    }
\caption{\textbf{Pre-training and fine-tuning hyperparameters.} We use the same hyperparameters for both diffusion and non-diffusion models.  $^*$: Batch size refers to effective batch size. $^\dag$:``R'' refers to sampling random starting points with cyclic rolling in time when loading waveforms. ``N'' refers to adding random noise to the spectrogram. $^\ddag$: ``BCE'' is binary cross entropy, and ``CE'' is cross entropy.}
\label{table:training_details}
\end{table*}

\section{FLOPS Analysis}\label{section:appendix_flop_analysis}

In \cref{table:per_module_flops}, we summarize the reduction in FLOPS on a per-module basis for each of our efficiency strategies. Efficiency metrics are measured relative to the standard MAViL model in \cref{table:mavil_vs_diffmavil}. Row R1 is our DiffMAViL model with no efficiency strategies. Here we observe that the video encoder FLOPS are lower than MAViL's. This is because, in MAViL, the first step after patchifying the input is to project all the patches onto the encoder space and then mask them; however, in DiffMAViL, we first mask patches and subsequently project only the visible patches onto the encoder space. This is so that we can later diffuse the masked patches before projecting them onto the decoder space. Moreover, we observe that the decoder FLOPS in R1 are slightly higher than those of MAViL; we attribute this to the fact that, in DiffMAViL, the diffused masked patches must first be projected onto the decoder embedding space prior to being processed by the decoder. In contrast, the standard video decoder without diffusion only projects visible patch embeddings output by the encoder since the \mask tokens are already of the appropriate dimension. In row R2 we observe that the use of cross-attention reduces the video decoder FLOPS by about 47\%. In row R3, we observe that the additional use of a linear masking ratio schedule reduces audio and video encoder FLOPS by about 26-28\%. This is because a higher masking ratio yields fewer visible patches, and therefore fewer patches are processed by the encoders.

\begin{table*}[h!]
    \centering
    \resizebox{\columnwidth}{!}{
    \begin{tabular}{ccc|ccccc|c}
        \toprule[1.5pt]
        \textbf{Row \#} & \textbf{Masking Ratio} & \textbf{Video Attention} &  \textbf{Audio Encoder} & \textbf{Audio Decoder} & \textbf{Video Encoder} & \textbf{Video Decoder}  & \textbf{Fusion Encoder} & \textbf{Total} \\
        \midrule[1.25pt]
        R1 & Fixed & self & 1.0 & 1.0 & 1.0 & 1.0 & 1.0 & $1.0\times$ \\
        R2 &  Fixed & cross & 1.0 & 1.0 & 0.97 & 0.53 & 1.0 & $0.81\times$ \\
        R3 & Linear & cross & 0.74 & 1.0 & 0.72 & 0.54 & 0.74 & $0.68\times$ \\
        \bottomrule[1.5pt]
    \end{tabular}
    }
\caption{\textbf{FLOPS reduction in audio/video encoders and decoders due to use of diffusion, cross-attention, and a masking ratio schedule.} The use of cross-attention instead of self-attention in the video decoder reduces total pre-training FLOPS by 19\%. Adding a linear masking ratio curriculum further reduces the pre-training FLOPS by 32\%. Efficiency metrics are reported relative to the standard MAViL model in \cref{table:mavil_vs_diffmavil}.}
\label{table:per_module_flops}
\end{table*}

\section{AudioMAE + Diffusion}\label{section:appendix_audio_mae_and_diffusion}
In this section, we consider integrating diffusion into the AudioMAE \cite{AudioMAE} framework. As described in \cref{section:diffmavil}, we simply replace the learnable \mask tokens in the decoder with diffused spectrogram patches. In \cref{table:diff_audiomae}, we compare the downstream performance of our implementation of AudioMAE with AudioMAE + Diffusion. We observe that training with diffusion improves performance in downstream tasks. 

\begin{table*}[h!]
    \centering
    \resizebox{0.7\columnwidth}{!}{
    \begin{tabular}{ccccc}
        \toprule[1.5pt]
        \textbf{Diffusion} & \makecell{\textbf{AS-20K} \\ \small{(mAP $\uparrow$)}} & \makecell{\textbf{VGGSound} \\ \small{(Top-1 $\uparrow$)}}  & \makecell{\textbf{ESC-50} \\ \small{(Top-1 $\uparrow$)}}   & \makecell{\textbf{SPC-v2} \\ \small{(Top-1 $\uparrow$)}} \\
        \midrule[1.25pt]
        \xmark & $34.2 \pm 0.06$ & $57.1 \pm 0.16$ & $92.6 \pm 0.12$ & $98.4 \pm 0.03$ \\
        \cmark & $35.5 \pm 0.07$ & $57.9 \pm 0.08$ & $93.6 \pm 0.02$ & $98.4 \pm 0.05$ \\
        \bottomrule[1.5pt]
    \end{tabular}
    }
\caption{\textbf{Diffusion improves the performance of AudioMAE.} We augment the AudioMAE \cite{AudioMAE} framework with diffusion and observe that diffusion facilitates the learning of richer audio representations in the absence of the video modality. Both models (with and without diffusion) were pre-trained on the AS-2M \cite{audioset} dataset.}
\label{table:diff_audiomae}
\end{table*}

\end{document}